# 3D Printing-Assisted Energy Loss Testing of Artificial Aortic Heart Valves

Yuqi Jin[1#], Akansha Rao[1#], Tae-Youl Choi, Ph.D [1*]
Mechanical and Energy Engineering
University of North Texas, Denton TX 76207, USA
*Corresponding author: Tel. 940-565-2198;
Email: tae-youl.choi@unt.edu

William T. Brinkman, MD[2]
The Heart Hospital Baylor Plano, Plano TX 75093, USA

*Abstract*—With development of additive manufacturing, the 3D printing technology has become more common nowadays. We designed and 3D printed a cost-effective liquid fluid pressure measurement device which can convert liquid pressure to air pressure. Air pressure was measured by a low-cost manometer in this study. The device can measure pressure drop across artificial aortic heart valves at different flow rate conditions. The main objective of this study is to utilize the 3D printing technology to measure pressure loss through artificial aortic heart valves. From the measured pressure drop, the energy loss of flow that occurs through the artificial aortic heart valves can be estimated. In the experiment, both mechanical and biological (bovine skin) aortic heart valves have been tested at several different flow rate conditions. By doing so, we were able to measure energy loss at different flow rates through the valves to test which type and size of heart valve has lowest energy loss during operation. . Lower energy loss of the blood through the valve should have lower burden to patient's heart. This study provides an example of a new application of additive manufacturing in liquid fluid pressure measurement device and showed an acceptable performance in experiment.

## I. INTRODUCTION

A heart consists of four main valves: the mitral valve, tricuspid valve, aortic semilunar valve and the pulmonary semilunar valve. The functions of the mitral and tricuspid valves are to control the blood flow from the atria to the ventricles[2]. The primary function of the aortic and pulmonary valves is to transport blood out of the ventricles. The aortic valve is located on the left side of the heart and is the gateway between the body and the heart[3]. It opens when the left ventricle squeezes to pump the blood out and then shuts close in the middle of heart beats to keep the blood flow from prolapsing.

Aortic valve diseases occur in different forms, namely aortic regurgitation and aortic stenosis[4]. Aortic regurgitation occurs when blood flow is going the opposite way, which is result of the aortic valve closing only partially[5]. Another disease, aortic stenosis, is when the valvar structure does not open completely, so the body is deprived of blood flow because the heart does not pump enough blood[6]. Some common causes and symptoms of aortic valve diseases can be congenital (detected from birth) or can be acquired as the patient becomes older.

Leaflet configuration is the arrangement of the leaf like flaps of muscle, which are primarily found in heart valves. All valves in a heart have different functions, but for the means of this paper, the aortic valve will be the focus[7]. The formation of the semilunar valves begins at four weeks from the beginning of the gestation period, which is when the heart also starts to beat for the first time[8, 9]. During this early stage in the development of the fetus, the formations of the valves are extremely fragile, and they will move around the structure of the heart. The aortic, pulmonary and tricuspid valves typically have a trileaflet structure, whereas the bicuspid valve has a bileaflet structure[10, 11].

Energy loss of an aortic valve typically occurs through the orifice area of the valve, which is the gap between the openings of the leaflets. Since the flow of blood is one continuous motion, the abruptions in the energy are found at the "stops", in this case, the various valves inside the heart. Through the aortic valve, the energy loss occurs once the oxygenated blood is pushed from the left atrium, to the left ventricle, through the aortic valve, to the aorta, and then through the rest of the body. This is because the kinetic energy of the blood moves upwards through the valve and to the aorta, resulting in a minute amount of energy loss of blood flow specifically through the aortic valve.

For the purposes of our experiment, 3D printing was used to fashion a pressure system as well as placement pieces for the aortic valves. By doing so, there was a limited amount of energy loss as well as proper structure for where the aortic valve was held into place, also assisting with minimizing the energy loss. With the use of 3D printing, a modern method was implied to complete the scientific analysis of energy loss through artificial aortic valves.

As some previous research and experiments have shown, the energy loss of aortic valve mainly is calculated from flow speed gradients[1]. Doppler flow speed measurement has been used which is costly. Moreover, the energy loss of aortic valves is not only from flow speed gradients but also from flow pressure drop. As the solution would flow through the valves, the energy loss and pressure drop were determined with the help of a low-cost manometer and flow meter. The device is capable of measuring various liquid pressure difference at different flow rate conditions with no flow speed difference between inlets and outlets of the valve. To measure the pressure difference between the flow before going through the valves and after going through the valves, a cost-effective liquid fluid pressure measurement device which can convert liquid pressure to air pressure was designed and 3D printed. Air pressure was measured by a low-cost manometer in this study. The main objective of this study is to design customized fluid pressure measurement devices that fit with various artificial aortic heart valves with irregular contours using the 3D printing technology and to measure pressure loss through artificial aortic heart valves. From the measured pressure drop, the head loss of the artificial aortic heart valves can be estimated to find energy loss. In the experiment, both mechanical and biological (bovine skin) aortic heart valves have been tested at several different flow rate conditions

## II. EXPERIMENT

To measure energy loss of biological and mechanical aortic valves, flow pressure difference between the inlet and outlet of the valves need to be measured. Fig. 1 shows a schematic diagram of the experimental setup. The experiment was performed with artificial heart valves, pump, water tank, flow meter, and manometer. The water flow rate from the water tank was measured by a liquid flow rate meter with 0.25 liter per min resolution. To maintain the flow velocity at constant level which could eliminate the dependency of flow rate on energy loss measurement, we have used a flow rate controller. The difference pressure was measured by air pressure manometer with a digital readout. The water pump can deliver up to 5.5 gallons per minute of flow rate.

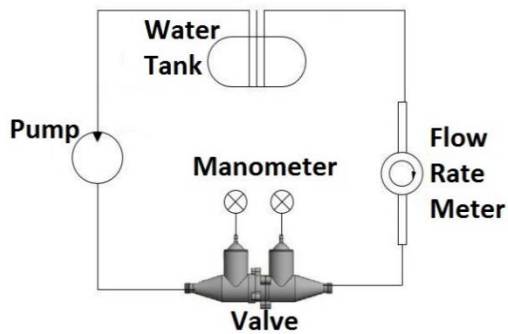

Figure 1 The Characterization of the Entire Experiment Instrument. Flow direction is counter-clock wise. Testing valve is in the middle place of 3D printed measurement device. Manometer measures the pressure drop in front of the valve and behind the valve.

### A. Experiment Materials

In this experiment, totally four aortic valves had been tested which are Edward Lifesciences Perimount Magna Aortic Valve ThermaFix Process Model 3000TFX (19 mm diameter), Edward Lifesciences Perimount RSR bioprosthesis Aortic Valve Model 2800 (25 mm diameter), Edward Lifesciences Perimount RSR bioprosthesis Aortic Valve Model 2800 (21 mm diameter), and Mechanical Aortic Bileaflet Valve (25 mm diameter).

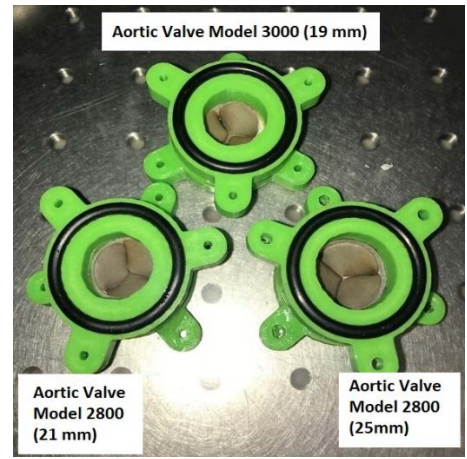

Figure 2 Three variable diameter Biological Valves Adhered between 3D Printed ABS Adaptors by epoxy with O-rings for testing

### B. 3D Printing Instrustments

Auxiliary parts to assist the pressure drop measurement across the valves were designed and 3D-printed. These auxiliary plastic adaptors were used to fit the specific needs for this experiment. Since the shape of biological valves is irregular, the interfacial contours of the biological valves needed to be designed by matching those of 3D printed adaptors. Each valve is sandwiched between and adhered with two pieces of printed adaptors by water-proofed epoxy, as Figure 2 and Figure 3 show. After adhesion, each specimen is connected to the 3D printed pressure measurement device as the location 1 showed in Figure 4. Between the connections, O-rings are used on each side to avoid leakage. As Figure 2 shows, the plastic adaptors were printed by ABS (acrylonitrile butadiene styrene) which has higher strength and relatively smooth surface. Not only valve adaptors, the whole pressure drop measurement device was designed and 3D printed by ABS.

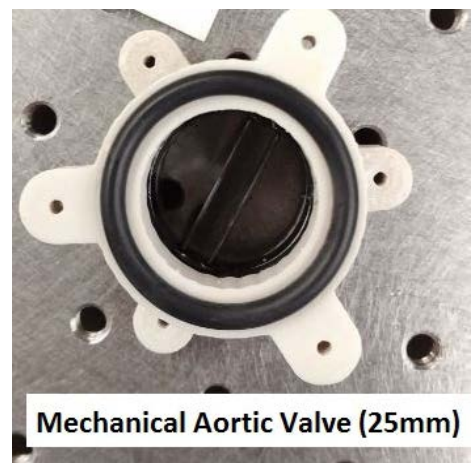

Figure 3 Three variable diameter Biological Valves Adhered between 3D Printed ABS Adaptors by epoxy with O-rings for testing

## C. Experiment Methods

The energy loss of the flow though the valves is mainly due to the flow pressure dropping. To measure the pressure drop between inlet and outlet of the valves, instead of using costly liquid pressure meter, the 3D printed pressure measurement device can convert liquid pressure to air pressure by compressing the air pockets with in the device (location 2 and 3 in Figure 4). Then, the air pressure is measured by a relatively low-cost manometer with 0.01 Pa resolution.

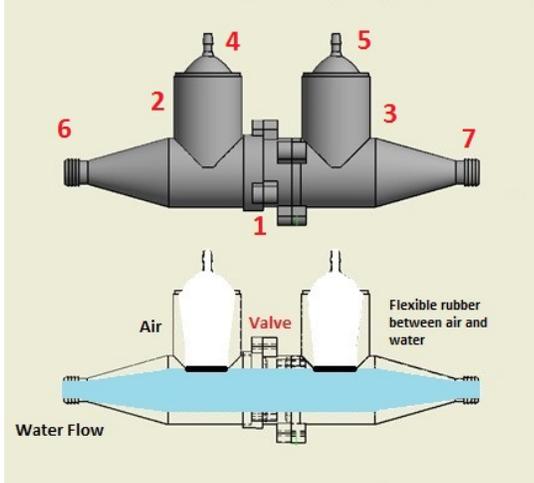

*Figure 4 Fluids Flow inside of 3D Printed Pressure Drop Measurement Device, and working principle of the device converting water pressure to air pressure. Flexible rubber between air and water showed in picture as two bold black line.*

As Figure 4 shows, the flow goes into this system from location 6 and comes out of the system from location 7. At locations 2 and 3 there are two air pockets which are in contact with water flow. Between the air and water flow, there is a piece of flexible rubber as Figure 4 shows. Through the locations 4 and 5 two pressure ports in the manometer (air pressure meter) are connected. Deformation of the rubbers due to fluid pressure increases the air pressure within the pockets. In this way, the water flow pressure could be converted to air pressure which could be measured by manometer. After the flow rate being close to steady state, the manometer measured the differential air pressure between the two sides of valves. It should be noted that the pressure drop without the valves installed (even though negligible) has been measured to accurately characterize the pressure drop across the valves; the pressure drop measured without the valves were subtracted from the one with the valves. The measurements of pressure drop from manometer are in the unit of psi which is converted to Pa in order to get energy loss in J/m$^3$. The flow rates were 2.5, 3.0, 3.5, 4.0, and 4.5 liters per minutes.

## III. RESULTS AND DISCUSSION

Energy Balance Equation for two locations (1 and 2) of fluid in pipe can be written with all the head (energy) losses involved.

$$\frac{P_1}{\rho g} + \frac{v_1^2}{2g} + Z_1 = \frac{P_2}{\rho g} + \frac{v_2^2}{2g} + Z_2 + h_f + \sum h_m \quad (1)$$

Where $h_f$ is head loss due to the friction of the pipe wall which can be obtained from the measurement without the valve installed and $h_m$ is the head loss due to the valve. $P$, $v$, $Z$, $\rho$, and $g$ are pressure, velocity, elevation, density, and gravity, respectively.

$$\sum h_m = \left(\frac{P_1}{\rho g} - \frac{P_2}{\rho g}\right) + \left(\frac{v_1^2}{2g} - \frac{v_2^2}{2g}\right) + (Z_1 - Z_2) - h_f \quad (2)$$

In the present experiment, there is no change in elevation and velocity. Therefore, the energy loss due to valves can be written:

$$(\sum h_m)\rho g = (P_1 - P_2) - h_f \rho g \quad (3)$$

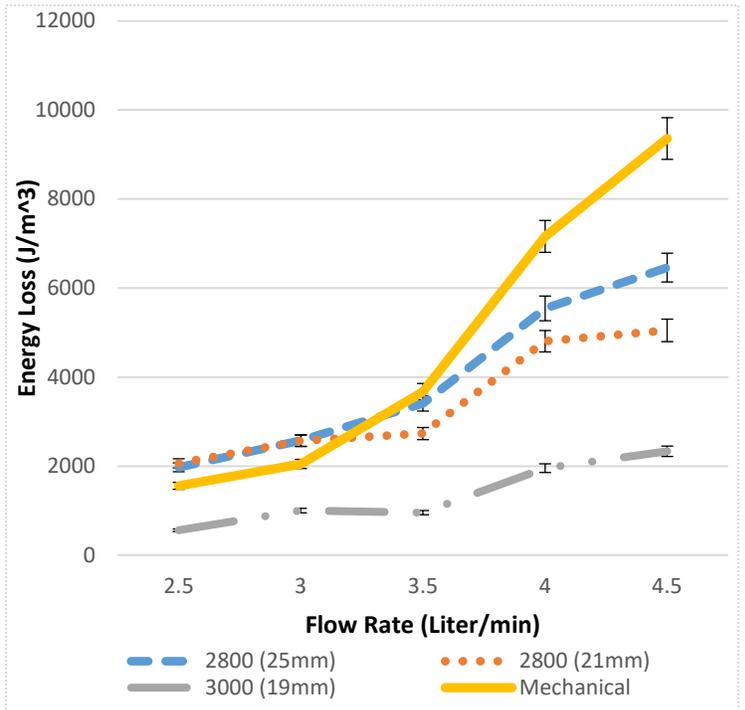

*Figure 5 Comparison plots in energy loss of each aortic valve*

The results were obtained for 5 trial pressure drop values of each valve, which correspond to volumetric flow rates of 2.5, 3, 3.5, 4 and 4.5 liter per minutes. The average pressure drop values are subtracted by average head loss values which were measured by the same instrument without heart valves. The mechanical valve has the highest energy loss at high flow rate conditions. Since the solid plastic leaflets of the mechanical valve are much thicker than the leaflets on the biological valves (the thickness of the leaflet on mechanical valve are more than 1.2 mm), when the high-speed flow passed, the mechanical leaflet created more friction comparing to the biological leaflets whose thickness of the leaflets are around 0.3 mm. Moreover, the biological leaflets are relatively much softer than mechanical leaflets, so friction generation through the biological valves is less than that of mechanical valve. At low flow rate conditions, however all the valves including the mechanical valve have similar amount of energy loss. The energy loss of 25 mm biological valve and the energy loss of 21 mm biological valve are very close. The energy loss of 25 mm biological valve is even slightly smaller than 21mm biological valve at 2.5 liter per min. The biological valve with a larger orifice area has more energy loss compared to the valve with a smaller orifice area at the flow rate above 3.5 liters per minute because the bigger biological valves have larger leaflets which produces more friction when flow passes through it. At 4.5 liters per min flow rate condition, the energy loss of the mechanical valve is around four times larger than Model 3000 19 mm biological valve. Therefore, it will become a greater burden for a patient who uses a valve with larger energy loss.

## IV.   CONCLUSION

To compare the performance of different aortic valves by measuring flow energy loss, energy loss is determined by flow pressure drop, when the flow passing through the aortic valves. To measure the pressure drop, a special device has been designed and 3D printed to convert the liquid pressure to air pressure. The 3D printing pressure measurement device has assisted this experiment well. The 3D printing instruments convert liquid pressure to air pressure which could be measured by a manometer. Moreover, 3D printing technology also help making customized (different size and irregular shape) aortic valve adaptors for connecting pressure measurement device and valves to achieve the goal of this study. The pressure difference values were measured by manometer. Since all aortic valves have different orifice areas and shapes, the adaptors for each valve were designed for the specific valves. The pressure drop values were measured from different flow rate conditions and averaged from 3 controlled condition trials. From the results of this experiment, energy loss of mechanical valve is higher than biological valves at higher flow rates. Comparing three biological valves, the bigger valve has larger energy loss. The probable reason is the bigger valve has bigger leaf which might cause larger flow resistance. Overall, the model 19mm biological valve has the lowest energy loss, which should have lowest burden to patient's heart.

## ACKNOWLEDGEMENT

This work was supported by the Korea Institute of Energy Technology Evaluation and Planning (KETEP) and the Ministry of Trade, Industry & Energy (MOTIE) of the Republic of Korea (No. 20168510011420).